\documentclass[lettersize,journal]{IEEEtran}
\usepackage{amsmath,amsfonts}
\usepackage{algorithmic}
\usepackage{algorithm}
\usepackage{array}
\usepackage[caption=false,font=normalsize,labelfont=sf,textfont=sf]{subfig}
\usepackage{textcomp}
\usepackage{stfloats}
\usepackage{url}
\usepackage{verbatim}
\usepackage{graphicx}
\usepackage{cite}
\usepackage{nccmath}
\usepackage{xcolor}
\usepackage{nccmath, mathtools}
\def\BibTeX{{\rm B\kern-.05em{\sc i\kern-.025em b}\kern-.08em
		T\kern-.1667em\lower.7ex\hbox{E}\kern-.125emX}}
\usepackage{balance}

\begin{document}

\title{Hardware-Efficient and Performance-Enhanced Joint Pulse Shaping and Dispersion Compensation for Coherent Data Center Interconnects}

\author{Yukun Zhang, Xiaoxue Gong, Weigang Hou, Xu Zhang, Lei Guo
    
\thanks{This work was supported in part by the National Key Research and Development Program of China under Grant 2023YFB2905900, in part by the National Natural Science Foundation of China under Grants U24B20134, 62222103, in part by the Chongqing Municipal Education Commission under Grants KJZD-K202400608.

Yukun Zhang, Xiaoxue Gong (Corresponding author), Weigang Hou, Lei Guo are with the School of Communications and Information Engineering, Chongqing University of Posts and Telecommunications, Chongqing 400065, China 
			(e-mail: gongxx@cqupt.edu.cn).
			
			 Yukun Zhang, Xiaoxue Gong, Weigang Hou, Xu Zhang, and Lei Guo are with the Institute of Intelligent Communications and Network Security, Chongqing University of Posts and Telecommunications, Chongqing 400065, China.

         Weigang Hou, Xu Zhang, and Lei Guo are with the School of Computer Science and Engineering, Northeastern University, Shenyang 110819, China.
}}

\maketitle

\begin{abstract}
With the explosion of data traffic triggered by 5G/6G and Generative artificial intelligence, coherent optical communication is moving towards higher baud rates and more complex modulation formats. This leads to a significant increase in the computational complexity and power consumption of digital signal processing (DSP) at the transmitter and receiver ends, especially in the chromatic dispersion(CD) Compensation and low roll-off shaping filter modules.
We propose a joint shaping filtering and CD compensation (JFS-CD) algorithm. This algorithm moves the CD compensation to the transmitter side and utilizes the characteristics of discrete fourier transform and the spectral features of shaping filtering for integrated processing.
Aiming at the high peak-to-average power ratio (PAPR) problem caused by chromatic dispersion pre-compensation, we propose a low-complexity square boundary clipping algorithm(SBC).
Simulation results show that, under the premise of maintaining unchanged performance, JFS-CD can reduce the real multiplication complexity by about 46$\%$. Meanwhile, benefiting from the suppression of the effects of system nonlinearity and receiver IQ imbalance, the joint JFS-CD and SBC scheme improves the Q-factor by about 0.3 dB in experiments compared to the traditional post-chromatic dispersion compensation scheme.
This research provides a highly potential transmitter DSP solution for next-generation low-power and high-performance data center interconnects (DCI). 
\end{abstract}

\begin{IEEEkeywords}
Coherent optical transmission
\end{IEEEkeywords}

\section{Introduction}
\IEEEPARstart{D}{riven} jointly by 5G/6G communication, large-scale cloud computing, ultra-high-definition video streaming, and generative artificial intelligence large models, global data traffic is experiencing explosive growth.
As the core foundation of information infrastructure, coherent optical communication systems are being forced to evolve towards higher rates, wider spectrum, and more complex modulation formats to support this massive demand.
However, this increase in signal rate directly leads to a sharp rise in the load of coherent digital signal processing (DSP) chips.

In DSP demodulation algorithms, the CD compensation occupies a large amount of power consumption and chip area.
Traditional frequency domain equalization (FDE) technology usually adopts the Overlap-Save method. This method highly relies on large-point Fast Fourier Transform (FFT) and inverse transform (IFFT).
In high-baud-rate transmission scenarios, FFT/IFFT operations not only consume huge logical resources, but also bring high dynamic power consumption.
In order to alleviate this difficult problem, researchers have carried out extensive research on low-complexity algorithms in recent years.
Reference \cite{ref2} proposed a Roots of Unity Equalizer. By approximating the filter coefficients as equally spaced phase shifts, it relies on shifts and adders to complete chromatic dispersion compensation.
Reference \cite{ref1} utilizes Chirp-Filtering technology.  While ensuring the compensation performance of high-order QAM signals, it reduces the computational complexity by nearly half.
These schemes reduce partial signal performance in exchange for the reduction of complexity.

On the other hand, the computational complexity of shaping filtering in high-baud-rate systems is also not to be ignored. 
In the time domain, a low roll-off factor will cause the impulse response of the filter to present an extremely slow attenuation characteristic.
Therefore, in order to maintain the high quality of the signal, the shaping filter needs to use a Finite Impulse Response (FIR) filter with dozens of taps.
A large number of taps brings more Multiply operations. This similarly increases the power consumption of the transmitter-side DSP.

In this paper, we propose a joint shaping filtering and CD Compensation (JFS-CD) algorithm to reduce the computational complexity required for shaping filtering and chromatic dispersion compensation.
In our scheme, the CD compensation module is moved to the transmitter side. It is combined with the shaping filtering to simultaneously achieve signal shaping filtering and CD compensation.
By utilizing the scale transformation characteristics of the Discrete Fourier Transform and the spectral features of shaping filtering, the computational complexity required for shaping filtering and CD compensation is reduced.
CD pre-compensation will lead to an increase in the PAPR of the waveform.  We propose a low-complexity square boundary clipping algorithm(SBC). Experimental/simulation results show that JFS-CD can achieve shaping filtering and CD compensation without performance loss.
The real multiplication complexity is reduced by about 46$\%$.
Moreover, in experiments, compared to the post-CD compensation scheme, the joint application of JFS-CD and SBC algorithms can improve the Q-factor by about 0.3 dB.
\section{Working principle}
\begin{figure}[t!]
\centering
\includegraphics[width=3in,keepaspectratio]{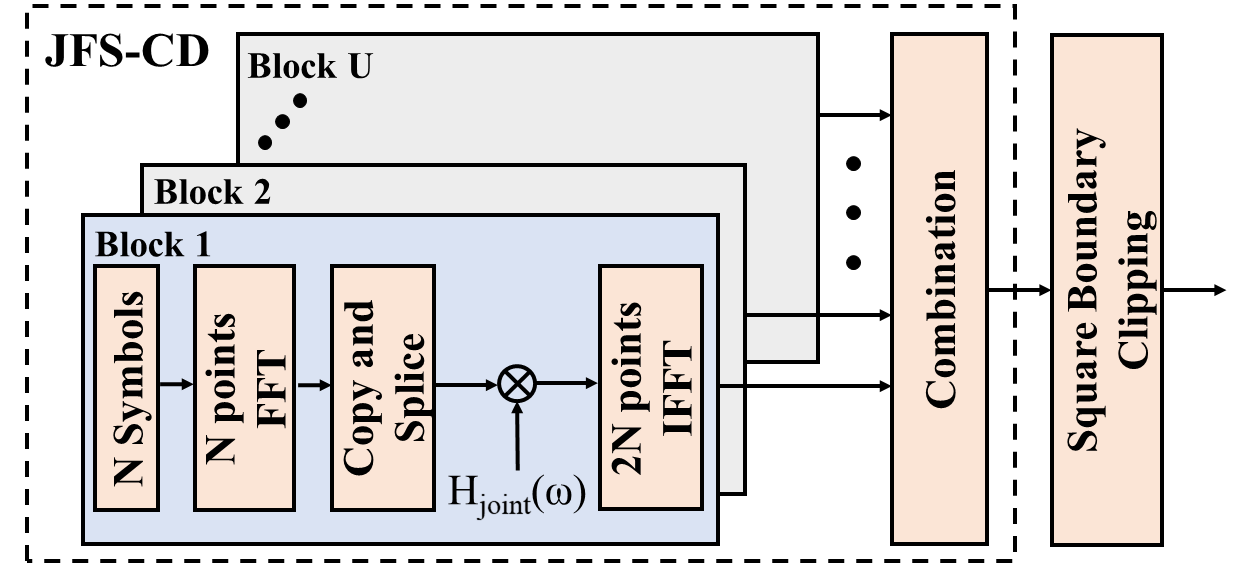} 
\captionsetup{justification=justified,singlelinecheck=false}
\caption{Schematic diagram of JFS-CD and SBC. }
\label{Fig.1}
\end{figure}
Figure \ref{Fig.1} shows the principles of JFS-CD and SBC.Assume that the time-domain impulse response of the pulse shaping filter of the system is $h(t)$. Its corresponding frequency-domain transfer function is $H_{p}(\omega)$.
The chromatic dispersion effect of the single-mode fiber can be modeled as an all-pass filter in the frequency domain. Its transfer function is
\begin{equation}
{H_{{\rm{CD}}}}(\omega ) = \exp \left( { - j\frac{{{\beta _2}}}{2}{\omega ^2}{\rm{L}}} \right)
\label{eq1}
\end{equation}
$\beta_{2}$ represents the chromatic dispersion coefficient. L represents the signal transmission length.The cascade of linear time-invariant systems is equivalent to the multiplication of frequency-domain responses. Therefore, the transfer functions of pulse shaping and CD pre-compensation are multiplied in the frequency domain to construct the simplified joint transfer function ${H_{joint}}(\omega ) = {H_p}(\omega ){H_{CD}}(\omega )$
\begin{equation}
H_{\text{joint}}(\omega) = 
\begin{cases} 
T_s  \exp \left( j\frac{\beta_2}{2}\omega^2 L \right), & |\omega| \le \frac{\pi (1 - \alpha)}{T_s} \\
0, & |\omega| > \frac{\pi (1 + \alpha)}{T_s} \\
C(\omega) \exp \left( j\frac{\beta_2}{2}\omega^2 L \right), & \text{otherwise}
\end{cases}
\label{eq2}
\end{equation}
\begin{equation}
C(\omega) = \frac{T_s}{2}\left[ 1 + \cos \left( \frac{T_s}{2\alpha} \left( |\omega| - \frac{\pi (1 - \alpha)}{T_s} \right) \right) \right]
\end{equation}
 $\alpha$ is the roll-off factor. $Ts$ is the symbol period. In the process of implementing JFS-CD, the transmitted symbol sequence is first divided into blocks. Each block contains N symbols. An N point FFT is performed on each block. According to the scale transformation property of the Discrete Fourier Transform (DFT), the N point frequency-domain data is periodically copied and concatenated according to the target oversampling rate. This generates a frequency-domain signal with a length of 2N. Then, this 2N length frequency-domain signal is multiplied point by point with Eq.\ref{eq2}.It should be pointed out that when $|\omega|>\pi(1+\alpha)/T_{s}$, Eq.\ref{eq2} is 0.
Therefore, the product of this part of the coefficients and the signal can directly be 0. This reduces the computational amount of multiplication.
Finally, a 2N point Inverse Fast Fourier Transform (IFFT) is performed on the multiplied frequency-domain signal. This generates an oversampled time-domain waveform.
In the IFFT stage, there are a large number of zero elements in the frequency-domain signal. Therefore, the computational complexity can be further reduced.
Assume that $X_{m}$ represents the time-domain signal of the $m$-th block after JFS-CD.
After completing the IFFT, the first $N_{o}$ points and the last $N_{o}$ points of the $X_{m}$ signal are polluted by circular convolution. Therefore, they need to be discarded.
The remaining data blocks also undergo the same operation. The final results are combined as the final output and sent into the SBC module.
Assume that the frequency-domain representation of the original baseband signal $x(t)$ is $X(\omega)=|X(\omega)|e^{j\theta(\omega)}$.
 $\theta(\omega)$ is the modulation phase jointly determined by the signal sequence and the shaping filter.
The transmitted signal after pre-compensation is s(t).
\begin{equation}
s(t) = \frac{1}{2\pi} \int_{-\infty}^{\infty} |X(\omega)| e^{j\left(\omega t + \theta(\omega) + \frac{\beta_2}{2}\omega^2 L\right)} \, d\omega
\label{eq3}
\end{equation}
Let it be $\Psi (\omega ,t) = \omega t + \theta (\omega ) + \frac{{{\beta _2}}}{2}{\omega ^2}L$. This integral term has a quadratic phase term. According to the principle of stationary phase, the main energy of this integral comes from the frequency band where the phase changes slowly with frequency. Let $
\partial \Psi (\omega ,t)/\partial \omega  = 0$
\begin{equation}
t =  - \frac{{\partial \theta (\omega )}}{{\partial \omega }} - {\beta _2}\omega L
\end{equation}

When a sufficiently long signal sequence is transmitted, some special data sequences will randomly exhibit a
specific quadratic phase distribution $\theta(\omega)$ in their spectrum. The derivative of this quadratic phase generates a linear term that cancels out the linear dispersion group delay $-\beta_{2}\omega L$, thereby eliminating the variable $\omega$. As a result, within a wide frequency range, it satisfies  $ - \frac{{\partial \theta (\omega )}}{{\partial \omega }} - {\beta _2}\omega L \approx {t_0}$. Eq.\ref{eq3} can be rewritten $s(t_{0}) \approx \frac{1}{2\pi}e^{jC} \int_{\Delta \omega } |X(\omega)|  \, d\omega$.
 At this time, ignoring the constant phase factor, the instantaneous signal amplitude is
\begin{equation}
|s({t_0})| \approx \frac{1}{{2\pi }}\int_{\Delta \omega } | X(\omega )|{\mkern 1mu} d\omega
\end{equation}
This means at time $t_{0}$, a high waveform peak is generated.
Therefore, there are two types of ideas to reduce the PAPR of the signal. One type is to destroy the formation conditions of phase consistency at the peaks, suppressing the generation of high peak values from the source. The other type is to clip the signal.
Due to the latter does not add extra system overhead.
Therefore, we propose a square boundary clipping algorithm suitable for CD pre-compensated signals. The principle of the SBC algorithm can be expressed as
\begin{equation}
y = \left\{ {\begin{array}{*{20}{l}}
{\frac{{x{A_{th}}}}{{\max (|Q\{ x\} |)}},}&{\max (|Q\{ x\} |) > {A_{th}}}\\
{x,}&{{\rm{otherwise}}}
\end{array}} \right.
\end{equation}
$A_{th}$ represents the threshold. $Q(\cdot)$ reprent real or imaginary parts.
When the waveform peak exceeds the threshold, the complex signal $x$ is scaled by the same  ratio. SBC performs simple threshold comparisons through the real and imaginary values of the signal.It avoids multiplication and square root operations. This reduces the computational complexity required for SBC.
\section{Simulation setup and Results}
\begin{figure}[t!]
\centering
\includegraphics[width=3in,keepaspectratio]{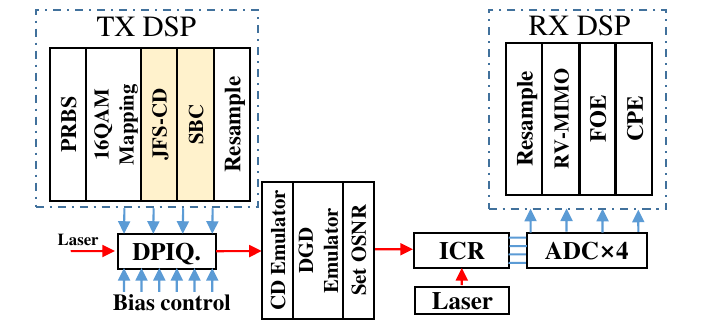} 
\captionsetup{justification=justified,singlelinecheck=false}
\caption{System setup and DSP flowchart }
\label{Fig.2}
\end{figure}
In order to study the performance of our proposed JFS-CD and SBC, we use 36 Gbaud signals with 2 samples per symbol(SPS) to perform simulations and experimental tests.
The system simulation setup and DSP flow are shown in Fig. \ref{Fig.2}.
The signal use a PRBS sequence with a length of $2^{18}$.
The ROF is 0.2.
The transmission laser and the LO laser are operating at 193.1 THz with a 100 kHz linewidth and a 1
GHz frequency offset.
The CD parameter is 16 $ps/(nm\cdot km).$
The transmission distance is set to 100 km.
A high-order PMD model is used to simulate the DGD effect.
The OSNR is defined assuming a bandwidth
of 12.5 GHz and is set to 23 dB.

\begin{figure}[t!]
\centering
\includegraphics[width=3in,keepaspectratio]{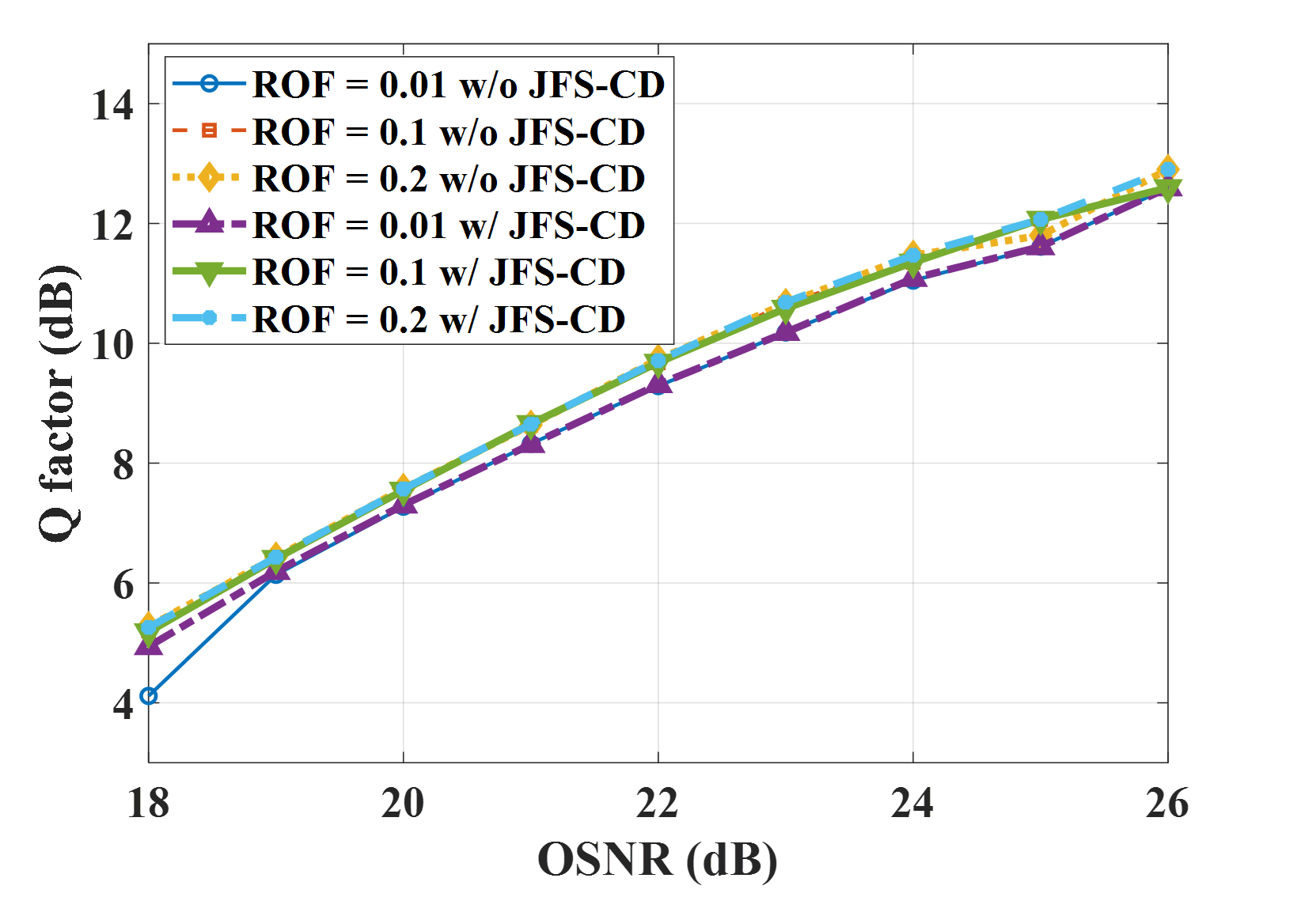} 
\captionsetup{justification=justified,singlelinecheck=false}
\caption{JFS-CD Performance Test Results}
\label{Fig.3}
\end{figure}
In order to verify the reliability of JFS-CD, Fig.\ref{Fig.3} evaluates the Q-factor performance of the system under different ROF. The evaluation compares the cascaded architecture (pulse Shaping + Pre CD compensation) and JFS-CD.
As shown in the figure, under ROF=0.01, 0.1, and 0.2, the performance curves of the traditional architecture and the JFS-CD architecture have a high degree of overlap.
Fig.\ref{Fig.3} shows that embedding the shaping filter into the frequency-domain chromatic dispersion operator can achieve a lossless equivalent transformation. It does not introduce any additional system penalty.

\begin{figure}[t!]
\centering
\includegraphics[width=3in,keepaspectratio]{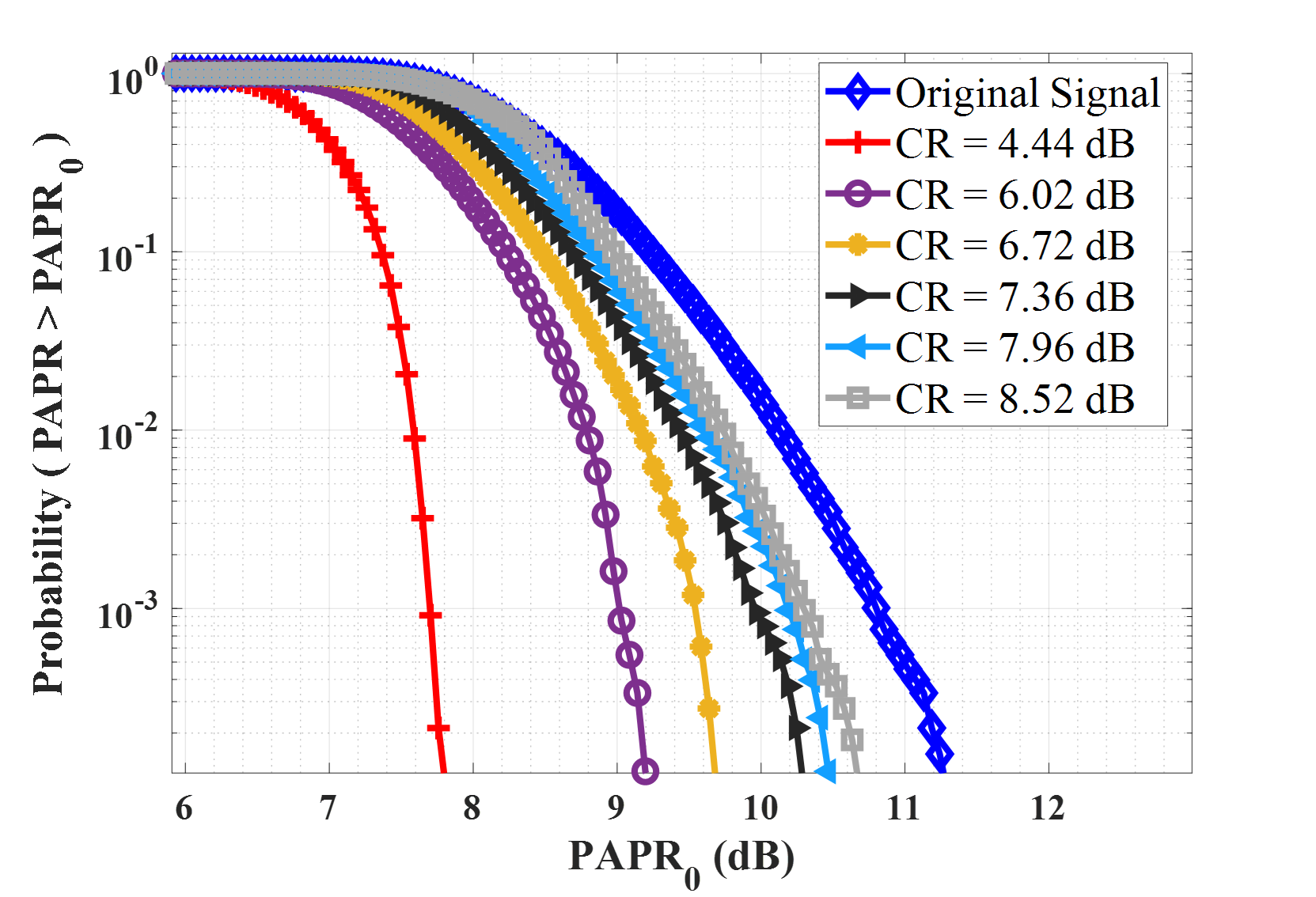} 
\captionsetup{justification=justified,singlelinecheck=false}
\caption{Test results of the CCDF performance curve for SBC}
\label{Fig.4}
\end{figure}
In Fig. \ref{Fig.4}, we use a PRBS sequence with a length of $2^{26}$ to conduct the CCDF performance curve study.
Clipping ratio (CR) is a relative threshold setting metric based on power normalization. 
The calculation of CR can be expressed as 
\begin{equation}
{\rm{CR (dB)}} = 20{\log _{10}}\left( {\frac{{{A_{th}}}}{{\sqrt {E[|x{|^2}]} }}} \right)
\end{equation}
As shown in Fig. \ref{Fig.4}, with the gradual decrease of CR, the CCDF curve presents a regular leftward-shifting trend. 
When CR=8.52 dB is adopted, it reduces the PAPR to 10.4 dB (about 0.5 dB gain) at a probability of $10^{-3}$.
When CR=6.72 dB is adopted, it obtains a peak suppression gain of 1.3 dB.

\begin{figure}[t!]
\centering
{\includegraphics[width=0.2412\textwidth]{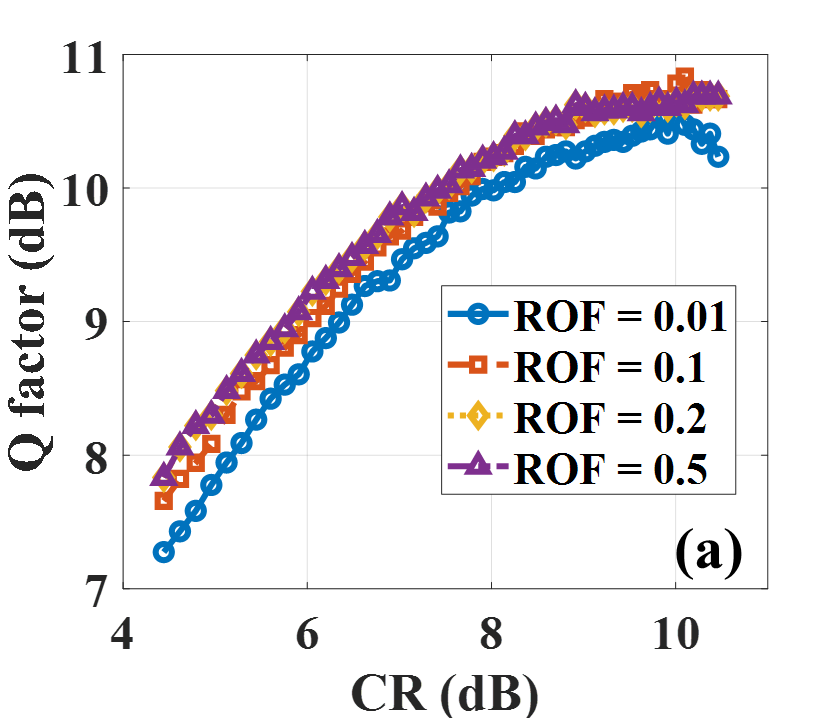}}
{\includegraphics[width=0.2412\textwidth]{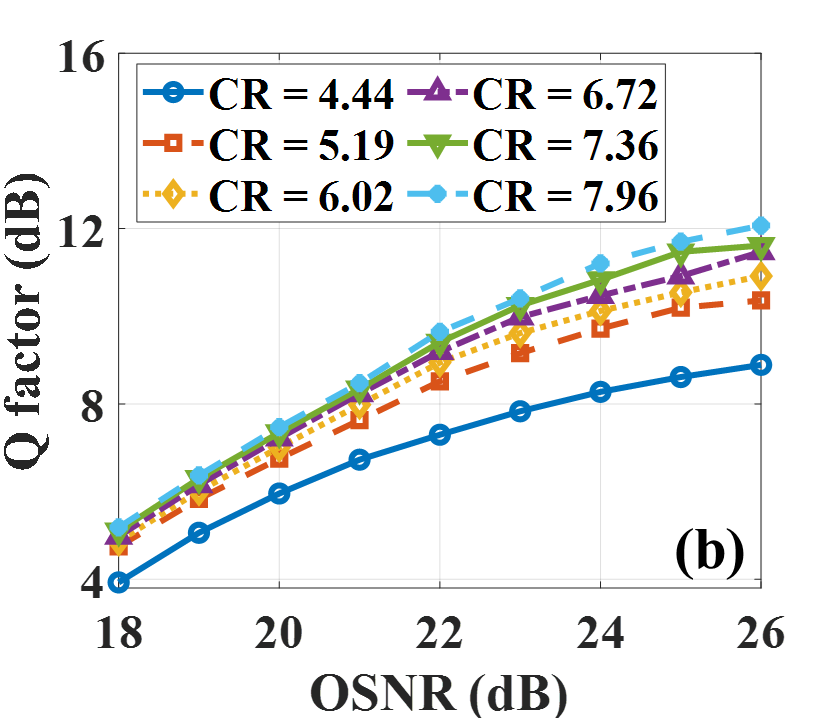}}
\caption{(a) CR performance curve test under different ROFs (b) System OSNR performance curve test under different CRs}
\label{Fig.5}
\end{figure}
In order to quantitatively evaluate the distortion cost introduced by SBC, Figure 5 shows the response curve of the SBC.
As shown in the Fig.\ref{Fig.5}(a), all curves present a highly consistent asymptotic behavior.
Fig.\ref{Fig.5}(a) shows the robustness exhibited by the SBC algorithm towards different ROF.
By comparing the attenuation slopes of different ROF curves, it can be found that the Q penalty introduced by SBC is not significantly amplified with the decrease of ROF.
In systems with low ROF, the signal easily generates peaks with huge amplitudes.
ecause the SBC algorithm is robust to ROF changes, using the SBC scheme can bring greater benefits after CD pre-compensation in low-ROF systems.

As shown in Fig.\ref{Fig.5}(b), the response curves of the  Q-factor varying with OSNR present significant interval differences under different CR.
In the low OSNR region ($<$21 dB), the distortion introduced by SBC is basically masked by the noise.
Therefore, the difference in the Q-factor penalty caused between different CR is relatively small.
However, as the OSNR gradually climbs to 26 dB, the limiting effect of noise is greatly weakened.
At this time, the in-band crosstalk caused by SBC becomes the bottleneck restricting signal quality.
Therefore, in the high OSNR condition, the gap of the Q-factor penalty caused by different CR expands accordingly.
Therefore, under low OSNR conditions, a low CR is used to reduce the PAPR to a greater extent.
Under high OSNR, a high CR is used to reduce the in-band crosstalk brought by SBC.

\section{Experimental setup and Results}
We investigate the performance of the proposed scheme through experiments.
The experimental setup is the same as the simulation setup shown in Fig.\ref{Fig.2}.
Two DFB lasers with a linewidth of $<$100 kHz and a wavelength of 1551.12 nm are used as the signal carrier and the local oscillator, respectively.
An Arbitrary Waveform Generator (Keysight M8196A) at 93.4 GSa/s is adopted to generate 36 GBaud DP-16QAM signals. The ROF is 0.01.
Subsequently, a dual-polarization optical IQ modulator (FTM7977)is used to modulate the signal onto the optical carrier.
The signal is transmitted in 100 km of standard single-mode fiber. The optical signal is amplified by erbium-doped fiber amplifiers (EDFA) at 0 km and 60 km, respectively.
At the receiver end, an Integrated Coherent Receiver (ICR, CPRV1225A) is adopted to demodulate the optical signal.
Finally, an oscilloscope (DSOZ594A) with a sampling rate of 80 GHz is used to sample the signal.
In the TX DSP, in order to prevent the resampled signal from entering the nonlinear region of the devices, it is necessary to perform normalization processing on the signal.

\begin{figure}[t!]
\centering
{\includegraphics[width=0.2412\textwidth]{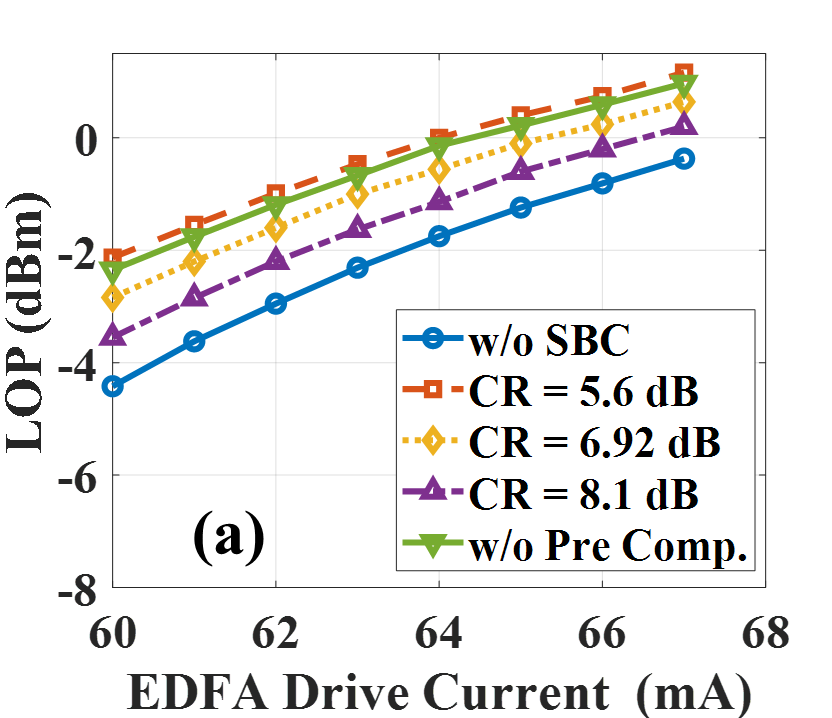}}
{\includegraphics[width=0.2412\textwidth]{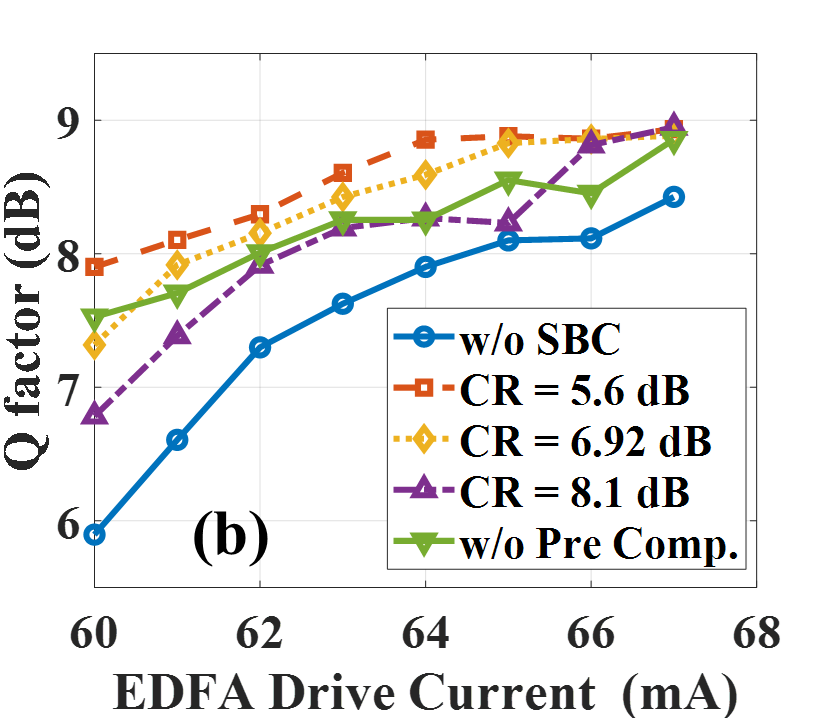}}
\caption{(a)  Measured launched optical power under different CRs (b) Measured Q factor under different CRs}
\label{Fig.6}
\end{figure}
Figure \ref{Fig.6} shows the relationship between the EDFA drive current and the actual launched optical power(LOP).
As shown in Fig. \ref{Fig.6}(a), compared to the signal without pre CD compensation(w/o pre comp.), the power of the CD pre-compensated signal without using SBC decreases by about 2 dBm.
After using the SBC algorithm, with the decrease of the CR, the LOP of the signal is significantly improved.
The improvement of power in Fig. \ref{Fig.6}(a) helps to improve the Q-factor of the signal.
Fig. \ref{Fig.6}(b) shows the performance of the Q-factor under different CRs.
For the signal without adopting SBC, its average power is too low. This causes its Q-factor to be at the lowest level throughout the entire testing interval.
Using SBC can effectively improve the Q-factor of the signal.
When CR=6.92 dB, the transmitted power is lower than that of the signal without pre-CD compensation. However, it shows a higher Q-factor quality than the signal without pre-CD compensation
There are two reasons for this. First, the signal without pre-CD compensation has a larger PAPR when it reaches the receiver. This brings a stronger nonlinear effect.
Second, because the signal use CD pre-compensation, the receiver IQ imbalance does not have a mixing effect with CD\cite{ref3}. The receiver IQ imbalance can be compensated by RV-MIMO.
The above two reasons jointly lead to the fact that the scheme proposed in this paper is better than the traditional post-CD compensation scheme in performance.

\section{Complexity Analysis}
We evaluate the computational complexity of the algorithm by the number of real multiplications required per symbol. When the ROF is 0.01 and N is 128 points (128 symbols).
JFS-CD requires a total of $8/N\left[ {N/2{{\log }_2}N + (1 + \alpha )N + N{{\log }_2}(2N)} \right]\approx101{\rm{ }}$ real multiplications.
In the shaping filter cascaded with CD compensation scheme, the shaping filter uses a 21-tap FIR filter. Each symbol requires $42 + 16{\log _2}(2N) + 16=186$ real multiplications. 
Our proposed JFS-CD reduces the computational complexity by about 46$\%$. When N = 64, the JFS-CD reduces the computational complexity by about 51$\%$.
 Due to a small number of symbols need to be adjusted using the SBC algorithm, the increased number of real multiplications can be ignored.

\section{Conclusion}
Aiming at the DSP power consumption and computational complexity bottlenecks faced by high-speed coherent optical communication systems, this paper proposes a JFS-CD algorithm.
Furthermore, it introduces a SBC algorithm to suppress the high PAPR caused by chromatic dispersion pre-compensation.
Research results show that the algorithm fusion is performed by utilizing the characteristics of Discrete Fourier Transform at the transmitter end.
Under the premise of not losing system performance, this scheme successfully reduces the real multiplication complexity by about 46$\%$.
Experiments further confirm that the joint application of JFS-CD and SBC can effectively suppress the effects of system nonlinearity and receiver-side IQ imbalance.
Compared to the traditional post-compensation scheme, it achieves a Q-factor improvement of about 0.3 dB.
In summary, while significantly reducing the hardware computational overhead, the scheme in this paper also takes into account the optimization of signal quality.
It provides a highly competitive DSP architecture reference for realizing next-generation low-power and high-performance coherent DCI.
\bibliographystyle{IEEEtran}
\bibliography{JFSCD_SBC}
\end{document}